\documentclass[twocolumn,aps,prl,superscriptaddress,showpacs]{revtex4}
\usepackage{amsmath,bm}%
\usepackage{graphicx}%

\begin{document}

%\draft
%\preprint{}
\title{ Sequential Decay Distortion of Goldhaber Model Widths for Spectator Fragments }
\author{Y. G. Ma}
\thanks{On leave from Shanghai Institute of Nuclear Research, Chinese
Academy of Sciences,  Shanghai 201800, CHINA}
\affiliation{Cyclotron Institute, Texas A\&M University, College
Station, Texas 77843-3366}

\author{R. Wada}
\affiliation{Cyclotron Institute, Texas A\&M University, College
Station, Texas 77843-3366}

\author{K. Hagel}
\affiliation{Cyclotron Institute, Texas A\&M University, College
Station, Texas 77843-3366}

\author{M. Murray}
\affiliation{Cyclotron Institute, Texas A\&M
University, College Station, Texas 77843-3366}

\author{A. Ono}
\affiliation{Department of Physics, Tohoku University, Sendai, 980-8578, Japan}

\author{J. S. Wang}
\thanks{On leave from Institute of Modern Physics, Chinese
Academy of Sciences, Lanzhou 730000, CHINA}
\affiliation{Cyclotron Institute, Texas A\&M University, College
Station, Texas 77843-3366}

\author{ L. J. Qin}
\affiliation{Cyclotron Institute, Texas A\&M University, College
Station, Texas 77843-3366}

\author{A. Makeev}
\affiliation{Cyclotron Institute, Texas A\&M University, College
Station, Texas 77843-3366}

\author{ P. Smith}
\affiliation{Cyclotron Institute, Texas A\&M University, College
Station, Texas 77843-3366}

\author{ J. B. Natowitz}
\affiliation{Cyclotron Institute, Texas A\&M University, College
Station, Texas 77843-3366}

%\email{}

%\address{}
\date{\today}

\begin{abstract}
Momentum widths of the primary fragments and observed final fragments 
have been investigated within the framework of 
an Antisymmetrized Molecular Dynamics transport model 
code (AMD-V) with a sequential decay afterburner (GEMINI). 
It is found that the secondary evaporation effects cause the values of a reduced 
momentum width, $\sigma_0$, derived from momentum widths of the final 
fragments to be significantly less than those appropriate to the primary fragment but
close to those observed in many experiments. Therefore, a new interpretation
for experiemental momentum widths of projectile-like fragments is presented.
 
\end{abstract}

\maketitle

Measurement of the momentum widths of projectile-like fragments (PLF) emitted in 
relatively peripheral nuclear collisions have long been viewed as a means 
of obtaining detailed information on the intrinsic Fermi momentum distribution 
in the nucleus 
\cite{Goldhaber,Viyogi,Greiner,Reinhold,Besliu,Bertsch,Murphy,Gan,Bauer,Brady,Dreute,MORRISSEY,GSI,Gossiaux}.
 This information is fundamental  in testing 
theoretical models of the nucleus and, in recent years, has taken on further 
importance in calculations related to the production of secondary fragmentation beams now 
being employed to study both structure and reactions far from stability \cite{Geissel}. 
Such studies, which explore a much broader range of neutron to proton asymmetry 
than previously accessible, are providing new insights into nuclear structure 
and nuclear astrophysics and offer the possibility of probing the 
nature of nucleonic matter in much greater detail \cite{see}.

In this Letter, we re-examine the validity of the Goldhaber Model \cite{Goldhaber}
within the framework of a non-equilibrium transport model, namely the AMD-V 
model of Ono et al \cite{Ono} and study the effects of sequential decay  on the observed 
momentum widths of  final-state spectator fragments with the help of an  
evaporation code GEMINI \cite{Charity}.  We find that the Goldhaber Model works well 
in describing the momentum widths of  the emerging primary fragments produced in  the 
spectator fragmentation. We find, further, that  reduced momentum widths derived from 
the observed  momentum widths of the  final fragments are narrower than the reduced 
widths characterizing the primary fragment momentum distributions. This reflects the 
influence of the light particle evaporation from  the primary fragments. Applying the 
Goldhaber Model without taking this evaporation  contribution into account leads to an 
erroneous interpretation regarding the primary fragmentation step.

In the pioneering work of Goldhaber \cite{Goldhaber}, it was suggested that the  
momentum widths of PLFs are determined by the intrinsic Fermi 
motion of the constituent nucleons which are  removed from projectile during the break-up 
process. Assuming that ($A_0-K$) nucleons are suddenly removed from a nucleus 
which originally has $A_0$ nucleons, a nucleus consisting of $K$ nucleons will emerge.
For this fragment of $K$ nucleons, Goldhaber showed that the momentum width, $\sigma$, 
could be expressed as :
\begin{equation}
\sigma = \sigma_0 \sqrt{ \frac{K(A_0 - K)}{A_0 - 1}}
\end{equation}
where  $\sigma_0$, the reduced momentum width, is related to the intrinsic Fermi motion 
of a single nucleon. If the projectile nucleons have a mean square momentum in the 
projectile frame equal to (3/5)$P_F^2$, where $P_F$ is the Fermi momentum, then  that a 
momentum dispersion with $\sigma_0^2$ = (1/5)$P_F^2$ is expected. Based upon 
electron scattering measurements of Fermi momenta \cite{Moniz} $\sigma_0$  
is expected to be $\simeq$ 112 - 116 MeV/c \cite{Goldhaber}. However,  experimental  
results  favor  values of $\sigma_0 \simeq$ 90 MeV/c \cite{Viyogi,Greiner,Reinhold,Besliu}.

In an attempt to explain this difference, Bertsch \cite{Bertsch} treated the correlations between 
the momenta of individual nucleons localized in space and calculated a corrected 
reduced width for $^{40}$Ar fragmentation which was 17${\%}$ smaller than 
Goldhaber's prediction. Murphy \cite{Murphy} considered the fact 
that the  fragment is also a Fermi gas, ie., that there also exists a phase space constraint on 
the nucleon momenta in the projectile fragments which was not taken into account in the 
Goldhaber Model \cite{Goldhaber}, and predicted fragment momentum distributions 
which are narrower than those observed. In a purely kinematical semi-classical model 
Gan et al. \cite{Gan} showed that the momentum width of a fragment suddenly broken off from a 
Fermi distribution is sensitive to the single-particle distribution  but the 
quantitative difference with the observed values still remains. Recently, Be\c{s}liu et al.
\cite{Besliu} took into account the dependence of $\sigma_0$ on the apparent temperature 
of the PLFs, which are excited during the collision. 
In this case, the projectile fragmentation process is assumed to be a slower process and 
the width governed by the Fermi distribution of nucleon momenta in the excited projectile 
as discussed by Bauer \cite{Bauer}. 
In contrast to most results for longitudinal momentum widths, 
Brady et al. observed  broader transverse momentum widths 
than those predicted  by the  Goldhaber Model \cite{Brady,Dreute},
These were explained by a collective motion: a bounce-off which imparts  an additional 
transverse momentum to the spectators and results in  wider momentum widths, 
especially for heavier projectiles.  

In recent years, several attempts have been made to provide a systematic phenomenological description 
of the available data \cite{MORRISSEY,Reinhold,GSI} and the  basic 
theoretical models have been extended \cite{Besliu}. In most of the theoretical work 
on the momentum widths of PLFs, analytical techniques have been 
employed  to explore the problem.  
One paper, that of Gossiaux et al. \cite{Gossiaux},  has employed a 
QMD  model calculation to explore the properties of spectator matter in 600 
MeV/nucleon collisions of Au with Au  and concluded that the underlying 
physics is more complicated than that generally assumed in analytical approaches.

The initial stage of the reaction studied in this paper, 200 MeV/u 
$^{40}$Ar + $^{27}$Al, has  been simulated using the AMD-V model \cite{Ono1}. 
AMD models have  
been very successfully used to study the static nature of light nuclei \cite{Horiuchi}. 
In this model,  
calculation of the minimum energy states can be carried out to define initial ground states 
of the colliding nuclei. 

In the  AMD-V model the stochastic branching process of the 
wave packet diffusion is incorporated. The widths of the wave packets in each stochastic 
branch are kept constant and the dynamics of the widths of the wave packets calculated 
by the Vlasov equation is incorporated as a stochastic diffusion process of the centroids 
of the wave packets. This model was successfully
applied to   multifragmentation events in the $^{40}$Ca + $^{40}$Ca reaction at 35 
MeV/u \cite{Wada} and $^{64}$Zn + $^{58}$Ni reactions \cite{Wada00} at 35 - 79 
MeV/u. In the present calculations, the  Gogny force \cite{Gogny}, which gave the best 
fit in these previous analyses, was used as an effective interaction. This Gogny force
gives a momentum-dependent mean field and an incompressibility of 228 MeV for 
infinite nuclear matter. We note further that in our initialization of the projectile and 
target ground states in the AMD-V model, we get an  initial width of the nucleon 
momentum distribution of  $\sim$ 105 MeV/c for $^{40}$Ar projectile 
and $\sim$ 101 MeV/c for the $^{27}$Al target.  
These values are slightly less than expected from  
electron scattering measurements of Fermi momenta, ie. 112 - 116 MeV/c \cite{Moniz},
but are still in the reasonable range.

The calculation was started at t =  0, with a distance of 15 fm between the
centers of the projectile and target in the beam direction. Each event was
followed up to t = 300 fm/c. However, considering that we are initially interested in the 
properties of primary fragments, we analyze the AMD-V results at an earlier time, when 
the two dominant spectators just separate from each other and before significant 
evaporation can occur.  
 
Calculations were carried out in the peripheral collision zone, 
ie. b = 6 - 8 fm. 10000 events are calculated.  Primary fragment masses  
and their excitation energies  were  extracted at $t_s$ = 60 $fm/c$.
Fragments were identified using a configuration space coalescence 
technique with a coalescence radius of 5 fm, but the size of the fragments depends only 
slightly on the coalescence radius.

The primary fragments which are isolated in this way are excited. In order to simulate  
experimental data as closely as possible, a modified version of the  GEMINI statistical model code 
\cite{Charity} has been used as an afterburner to follow the de-excitation of these  excited 
fragments. In this modified version, discrete levels of the excited states of light 
fragments with $Z \leq 14$ are taken into account and the Hauser-Feshbach
formalism is extended to treat the particle decay from a parent nucleus up to 50 MeV 
excitation energy.
In the calculation, care was taken to  follow the entire de-excitation cascade so that  the 
distribution of initial parent nuclei (primary fragments from AMD-V) leading to each  
observed final fragment could be derived and  the contributions of nucleon evaporation 
and mass loss to the momentum widths of final fragments could be determined. 

In order to study the primary fragments of spectator fragmentation,   
we  first defined, in each event, the heaviest fragment with a positive parallel velocity in the 
center of mass  as a projectile-like fragment  and the heaviest fragment with a 
negative parallel velocity in the  center of mass as a target-like fragment (TLF). We then analyzed 
the three components of the  momenta of the PLF and TLF fragments. Since the    
the collisions are peripheral, the primary PLF and TLF can be cleanly separated.

\begin{figure}
\includegraphics[scale=0.35]{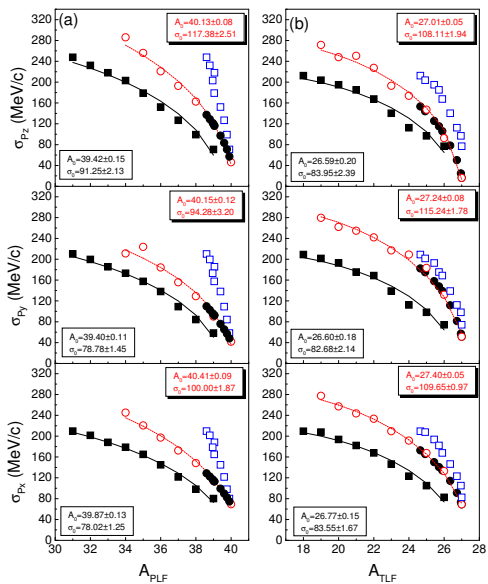}  
\caption{\footnotesize The three momentum widths $\sigma_{Pz}$
(upper), $\sigma_{Py}$ (middle) and $\sigma_{Px}$ (lower) as a function of the
mass of PLF (left) and TLF (right) fragments.   The open circles depict the momentum widths of the AMD-V 
primary fragments as a function of the mass of primary fragments and the
dashed lines indicate  the Goldhaber Model fits. The solid squares show the momentum widths of
final fragments as a function of the mass of final fragment 
and the solid lines represent Goldhaber Model fits to those results. The open squares represent 
the momentum widths of final fragments as a function of the reconstructed 
average mass of the primary fragment. The solid circles depict the reconstructed 
momentum widths as a function of the reconstructed average mass of the primary fragment. 
See details in text.}
\label{width_b68}
\end{figure}

These primary fragment  momentum distributions can be 
described by a Gaussian shape characterized by a width $\sigma$.  The primary fragment mass 
numbers and the three momentum widths, $\sigma_{Px}$, $\sigma_{Py}$ and $\sigma_{Pz}$ 
($Z$ is in the  beam direction and $X$ is in the direction of the impact parameter axis) 
are plotted in  Fig.~\ref{width_b68}  as a function of the  PLF and 
TLF fragment masses at  b = 6-8 fm. 
The widths of the primary fragments are denoted by  
open circles and the dashed lines represent the results of fits to the Goldhaber Model 
expression (Eq.(1)).   
In order to obtain better fits, we allow both the initial mass $A_0$ and 
the reduced width, $\sigma_0$ to be free parameters. 
The fit parameters  are shown in upper-right corner.   
The initial masses are close to the original masses of projectile or target
and the average of the reduced widths of the 
three momentum components is $\sim$ 105 MeV/c for PLFs and 110 MeV/c 
for TLFs. The values  of  $\sigma_0$ are very close to the initial nucleon 
Fermi momentum widths of the AMD-V ground states as mentioned before. 
Consequently, it appears  that the Goldhaber model works well for the primary fragments.

However, the calculation indicates that, the primary fragments are excited:
the mean value of excitation energy is about 1.7 MeV/u and 2.7 MeV/u for PLFs and TLFs,
respectively.   
To evaluate the effect of secondary particle emission on the masses and momentum widths 
 we used the AMD-V results at $t_s$ as input to the GEMINI calculation. As a result of the 
subsequent statistical de-excitation the final mass distribution becomes broader but  
the PLF and TLF components can still be identified and the momentum
widths for PLF and TLF final fragments can be determined.

The solid squares in Fig.~\ref{width_b68} 
represent the momentum widths of the final fragments, 
observed after the secondary de-excitation. 
The results have also been fit with the  
Goldhaber Model expression   (the solid line) and the extracted apparent primary mass $A_0$ and 
width $\sigma_0$ is shown on lower-left corner. 
The mass parameters  remain  close to the initial 
masses of the projectile and target but the $\sigma_0$ values  become significantly  
smaller,  manifesting mean values $\sim$ 83 MeV/c.
Obviously, these values approach to the typical 
values of 90 MeV/c observed experimentally (recall that the AMD-V ground state has a slightly lower 
Fermi momentum than those from experiments). It appears that the sequential 
decay plays an important role in causing the reduced  momentum width  
of the final fragments to be narrower than that of  the intrinsic nucleon  
momentum width. This leads us a relatively simple  explanation for the 
experimentally reported  momentum widths of PLF fragments.  

\begin{figure}
\includegraphics[scale=0.35]{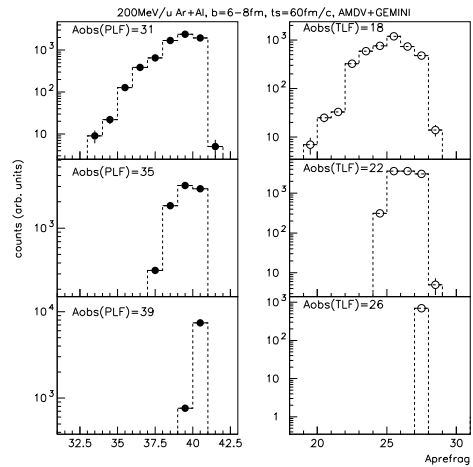}
\caption{\footnotesize The primary fragment  mass distributions for some
selected
PLF and TLF final fragments for 200 MeV/u $^{40}$Ar + $^{27}$Al at 6 - 8 fm.}
\label{Aini}
\end{figure}

Why does the sequential decay make the reduced width decrease and 
not increase? In order to answer this question, we have tracked the secondary decay 
paths and obtained the distribution of parent nuclei (primary fragments) which lead to  
each of the observed final fragments.  Fig.~\ref{Aini} shows some samples of 
the primary fragment distributions 
which contribute to some selected final fragments. Clearly, the farther removed the  final 
fragment from the primary parent  nuclei,  the broader the  distribution of primary 
fragments which contribute to that final  fragment. 

If we plot the observed momentum width for the final fragments as 
a function of their average primary fragment masses, as represented by the open squares in 
Fig.~\ref{width_b68}, %and Fig.~\ref{width_b46}, 
we immediately see that there is a large increase of momentum width in 
comparison with the original primary fragment width, denoted by open circles. 
It is, of course, natural that the sequential decay will make the momentum 
distribution wider.  

Knowing the parentage of a final fragment we can  reconstruct the primary width 
distribution leading to that fragment.  The reconstructed width for the average parent 
fragments can be written as:
$\sigma_{recons} \equiv \sqrt{\Sigma_i p_i  \sigma_{Prim}(i)^2}$, 
where $p_i$ is the fractional  contribution of primary fragments to the final observed  
fragments. The results are indicated by the solid circles in Fig.~\ref{width_b68}. 
As required by this procedure, the results fall on the curves defining 
the primary fragment momentum widths. The difference in widths represented 
by these points and those represented by the 
open squares reflect evaporation effects.

\begin{figure}
\includegraphics[scale=0.20]{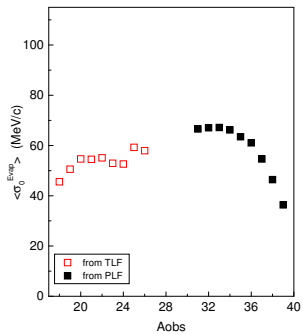}
\caption{\footnotesize Mean reduced momentum widths for light
particle emission from PLF and TLF primary fragments 
for 200 MeV/u $^{40}$Ar + $^{27}$Al at 6 - 8 fm.} 
\label{width_evap1}
\end{figure}

If we pursue the analogy to the Goldhaber Model, 
then from the decay calculations we can also define  
$\sigma_0^{Evap}$, a reduced width. This width characterizes the evaporation 
step and can readily be compared with the momentum width which results from the 
intrinsic nucleon  momentum distribution from the primary step. We  
define the  relationship between the width of the momentum distribution of final 
fragments of mass $A_{obs}$  and those of the primary fragments $A_{Prim}(i)$ 
which de-excite to produce $A_{obs}$ :
\begin{equation}
 \sigma_{Aobs}^2 = \Sigma_i p_i [ \sigma_{Prim}^2 + (A_{Prim}(i) - A_{obs})({\sigma_0^{Evap}})^2)]
\end{equation}
where $p_i$ is the fractional  contribution of $A_{Prim}(i)$  to the final observed  
fragments ($A_{obs}$)  and $\sigma_0^{Evap}$ is the contribution of particle evaporation
to the momentum width. 
Fig.~\ref{width_evap1} presents the mean value of  $\sigma_0^{Evap}$ over Px, Py and Pz 
directions as a function of the final fragment mass. 
We see that $\sigma_0^{Evap}$   is generally  in the range of 30-70 MeV/c, 
significantly lower than $P_F/\sqrt{5}$ which characterize the primary break-up.

For comparison, we also checked $^{40}$Ar + $^{27}$Al at b = 4 - 6 fm, the similar values
and the same conclusion was drawn. In addition, we also checked our technique to extract
$\sigma_0$ with the fixed $A_0$ as is typically 
done in experimental analysis of initial projectile and target masses in the fit procedure
with Goldhaber Model rather than the free parameter of $A_0$. The conclusions are not changed. 

In summary, the momentum widths of the primary fragments and final observed  fragments produced 
in the reaction  200 MeV/nucleon $^{40}$Ar + $^{27}$Al were investigated using an 
AMD-V calculation with a statistical model afterburner.   
It is found that the Goldhaber Model works well for the primary fragments
formed in spectator fragmentation. The momentum widths of the primary fragments  
are basically related to the initial Fermi momenta in the  projectile or target. 
However, since the primary fragments are excited, they are de-excited by  light 
particle evaporation.   This secondary  decay decreases   
the observed  masses and increases the observed  momentum widths of the primary 
fragments. As a consquence, it makes the  reduced width parameter $\sigma_0$,
 derived  from a Goldhaber 
Model fit to the data, narrower by almost 20 MeV/c than the initial 
Fermi momentum width, consistent with many experimental observations. 
Given that evaporation components can be experimentally determined from particle-
fragment correlation measurements, even in very complex reactions \cite{Marie},  
this result suggests that observed width distributions could be corrected for 
secondary decay and the Fermi momentum distribution of the primary fragment 
could be probed in relatively low intensity radioactive beam experiments. 
This would allow the extension of such measurements over a much larger range 
in N/Z ratio and  provide even more stringent tests of  our microscopic 
models of asymmetric nuclear matter.

This work was supported by the the U.S. Department of Energy and the Robert A. 
Welch Foundation.

{}

\end{document}